\documentclass{ws-procs9x6}
\usepackage{epsfig}

\def\CC{\mathcal{C}}
\def\CN{\mathcal{N}}
\def\p{\partial}
\def\hthet{\hat{\theta}}
\def\bari{\bar{i}}
\def\barj{\bar{j}}

\def\barz{\bar{z}}
\def\be{\begin{equation}}
\def\ee{\end{equation}}
\def\tr{{\rm tr}}

\def\CS{{\mathcal S}}
\def\CM{{\mathcal M}}
\def\del{\partial}
\def\pii{\pi^{-1}}

\begin{document}
%%%%%%%%%%%%%%%%%%%%%%%%%%%%%%%%%%%%%%%%%%%%%%%%%%%%%%%%%%%%%% 
% title, author(s) and address(es) put here:                 %
%%%%%%%%%%%%%%%%%%%%%%%%%%%%%%%%%%%%%%%%%%%%%%%%%%%%%%%%%%%%%% 
%\rightline{BRX-TH-531}

\title{
\vspace{-5\baselineskip}
\begingroup
\footnotesize\normalfont\raggedleft
\lowercase{\sf hep-th/0401233} \\ 
BRX-TH-531\\
PUTP-2109\\
\vspace{\baselineskip}
\endgroup
Remarks on branes, fluxes, and soft SUSY breaking\footnote{
\uppercase{T}o appear in the \uppercase{P}roceedings of the 3rd 
\uppercase{S}ymposium on \uppercase{Q}uantum
\uppercase{T}heory and \uppercase{S}ymmetries (\uppercase{QTS}3),
\uppercase{C}incinnati, \uppercase{O}hio,
10--14 \uppercase{S}ept 2003 ---
\copyright\
\uppercase{W}orld \uppercase{S}cientific.}}
\author{ALBION LAWRENCE}
\address{Physics Dept., Brandeis University,
Waltham, MA 02454\\
email: {\tt albion@brandeis.edu}}

\author{JOHN M\lowercase{c}GREEVY}
\address{Physics Dept., Princeton University, Princeton, NJ 08544\\
email: {\tt mcgreevy@princeton.edu}}


\maketitle

\abstracts{
We review recent work identifying 
soft SUSY-breaking terms
in local type II string models 
with branes and magnetic fluxes.
We then
make a new observation about the 
configuration space
of D-branes
in Calabi-Yau backgrounds,
and identify vevs for nonperturbative 
charged hypermultiplets in Calabi-Yau backgrounds
with $N=2$ Fayet-Iliopoulos terms.
}

%%%%%%%%%%%%%%%%%%%%%%%%%%%%%%%%%%%%%%%%%%%%%%%%%%%%%%%%%%%%%
% The main text of your paper                               %
%%%%%%%%%%%%%%%%%%%%%%%%%%%%%%%%%%%%%%%%%%%%%%%%%%%%%%%%%%%%%

\section{Introduction}

A wide class of phenomenologically attractive string theory 
backgrounds with low-energy $\CN=1$ SUSY are described 
by combinations of D-branes, orientifold
planes, and magnetic fluxes.  
%In such models 
Nontrivial gauge dynamics is typically localized in regions of 
the compactification manifold, and a fairly
generic scenario for SUSY breaking will
have supersymmetry broken in one region of the manifold,
with the standard model dynamics localized somewhere else.
Supersymmetry breaking will be communicated
via 10d supergravity effects at tree level, 
and via radiative corrections as 
in anomaly or gaugino mediation.
However, the detailed appearance and origins of such terms
in the low energy effective 
action of specific models is understood only 
in a few very specific examples.

In these proceedings we review and slightly extend 
recent work 
\cite{Lawrence:2004zk}\ studying 
the appearance of tree-level soft SUSY-breaking
terms for local models of Calabi-Yau threefold backgrounds
with D-branes. 
%and nontrivial geometric and topological features.  
The closed string modes
live in multiplets of $\CN=2$, $d=4$ SUSY. D-branes
and magnetic fluxes break SUSY to $\CN=1$ or $\CN=0$.
The D-brane modes, and some closed string modes controlling
the local geometry, have finite 4d kinetic terms even
when the CY is noncompact.  The other
closed string modes appear as ``spurions'',
as their dynamics decouples from the
low energy physics of the local model.  The auxiliary
fields for these modes appear as soft SUSY-breaking couplings
in the local model.  This provides a set of building blocks
for more complete models, and in a given model should allow
one to address questions such as whether the squark masses
are aligned with the quark masses.  Therefore, in Section 2 we
review the identification of auxiliary fields for light closed string
modes with magnetic fluxes, and review how these
appear as soft SUSY-breaking couplings on D-branes.
This section is based on the talk given by the first author at the
Quantum Theory and Symmetries 3 conference at the
University of Cincinnati.

Magnetic fluxes and D-branes are crucial aspects of type I,
type II, and F-theory models.  The interplay between these
two aspects of string theory makes apparent
some features of the string theory models which are
highly nontrivial from the point of view
of the low energy field theory.  In section 3 we discuss the
impact of fluxes on the space of D-branes in
Calabi-Yau compactifications.  In section 4
we discuss the degrees of freedom responsible
for tuning $\CN=2$ Fayet-Iliopoulos terms in
type II models on CY threefolds.

Because of the limited space allowed for these proceedings,
we will be 
%radically 
minimalist about referencing.  
A more complete bibliography 
appears in \cite{Lawrence:2004zk}.
We apologize to those who are not referenced here.

\section{Closed string modes and soft SUSY breaking}

\subsection{Auxiliary fields for $\CN=2$ vector multiplets}

We begin with type II string theory on
a local ({\it e.g.} noncompact) Calabi-Yau threefold $X$ times
${\mathcal R}^4$, together with D-branes filling 
${\mathcal R}^4\times \CC$, $\CC \subset X$.
Although the D-branes preserve at most $\CN=1$ supersymmetry
at most, the closed string modes lie in $\CN=2$ supermultiplets.
The $\CN=2$ properties of the latter constrain their couplings 
to the D-branes \cite{Brunner:1999jq}.  Therefore, it is
important to understand the underlying $\CN=2$ SUSY structure.
Vevs for auxiliary fields in the closed string supermultiplets
can break SUSY to $\CN=1$ or $\CN=0$ via computable operators
at tree level.

Closed string vector multiplets arise from complex structure
deformations of $X$.  We can write them as chiral superfields in
terms of the $\CN=2$ superspace variables $\theta, \hat{\theta}$
which are a doublet of Weyl spinors under the $SU(2)_R$ symmetry
of $\CN=2$ theories.  Translations in these superspace
directions are generated by spacetime supercharges
built from left- and right-moving worldsheet sectors, respectively.

A vectormultiplet can be 
described by a superfield $V$ which solves 
the chiral constraints
\begin{eqnarray}
\bar{\nabla}_{\dot{\alpha}} V & \equiv 
   \left(- \frac{\p}{\p\bar{\theta}^{\dot{\alpha}}} 
      - i \sigma^\mu_{\beta \dot{\alpha}}
      \theta^{\beta} \p_{\mu}\right) V = 0 \nonumber \\
   \hat{\bar{\nabla}}_{\dot{\alpha}} V & \equiv \left(- 
      \frac{\p}{\p\hat{\bar{\theta}}^{\dot{\alpha}}} 
      - i \sigma^\mu_{\beta \dot{\alpha}}
      \hat{\theta}^{\beta} \p_{\mu}\right) V = 0\ .
\end{eqnarray}
The superspace expansion for such a field is:
\begin{eqnarray}
   V & ~=~ w^a + \theta^\alpha 
   \zeta^a_\alpha+\hthet^\alpha\hat{\zeta}^a_\alpha
      + \theta^2 D_{++}^a + 
      \theta^\alpha\hthet^\beta
      \left(\epsilon_{\alpha\beta}D_{+-}^a + F^a_{\alpha\beta}\right)
      \nonumber\\
       & \ \ + \hthet^2 D^a_{--}
      + \theta^\alpha\hthet^2 \chi^a_{\alpha} +
      \hthet^\beta\theta^2
      \hat{\chi}^a_{\beta}\nonumber\\
      &\ \ \ + \theta^2 \hthet^2 C^a\ .
\end{eqnarray}
One may impose the additional constraints:~\cite{Lawrence:2004zk,deRoo:mm}
\be
(\epsilon_{ij}\nabla^i \sigma_{\mu\nu} \nabla^j)
   (\epsilon_{kl}\nabla^k \sigma^{\mu\nu} \nabla^l) V
   = - 96 \p^2 \bar{V}
\ee
which render $C$,$\chi$ as dependent variables; impose
the constraint that $\sigma^{\mu\nu}_{\alpha\beta}F^{\alpha\beta}$
be an anti-delf-dual tensor; and impose the ``reality
constraints'' $\Box D_{++} = \Box D_{--}^*$, $\Box D_{+-}$ real.

In type IIB string theory, we can identify the
bosonic degrees of freedom as follows.  If we
choose ``CFT coordinates'' on the moduli space
\cite{Lawrence:2004zk}, the scalar component $w^a$ can 
be associated to the perturbation
\be
\delta^m (ds)^2=\delta g^m_{\bari\barj}d\barz^{\bari}d\barz^{\barj}\ .
\ee
The label $m$ denotes a direction in the complex 
structure moduli space.  Factoring out reparameterizations,
each such deformation can be associated to an elements
of $H^{(2,1)}(X)$:
\be
   \omega^m_{\bari jk} = \delta g^m_{\bari\barj} g^{i\barj}\Omega_{ijk}
\ee
where $\Omega$ is the holomorphic $(3,0)$ form on $X$ and $g$ is the
metric.  Choose a basis $\omega^a$ 
of harmonic representatives of $H^{(2,1)}(X)$.
Auxiliary fields correspond to deformations of the NS-NS 3-form
$H = \sum_m h^m \omega^m + {\rm h.c.}$; of the RR 3-form
$F = \sum_m f^m \omega^m + {\rm h.c.}$; and of
$T = i (\p - \bar{\p})J = \sum_m \tau^m \omega^m + {\rm h.c.}$,
where $J = g_{i\barj}dz^id\barz^{\barj}$.  In terms of
components $h,f,\tau$:
\begin{eqnarray}
   D_{++}^m & = \left(\tau^m + h^m\right) \nonumber\\
   D_{+-}^m & = g_s (f^m - C^{(0)} h^m)\nonumber\\
   D_{--}^m & = \left(\tau^m - h^m\right)\ ,
\end{eqnarray}
where $C^{(0)}$ is the type IIB RR axion.

The results above can be proven 
using RNS worldsheet techniques.  Using these,
one may also find the auxiliary fields for
the ``special geometry'' coordinates on complex
structure moduli space. That is, choose a symplectic
basis $A^a,B_a$ of $H_3(X)$, such that
$A^a \cup B_b = \delta^a_b$, $A^a\cup A^b = B_a\cup B_b = 0$.
A good set of coordinates on moduli space
is $t^a = \int_{A^a} \Omega$.  The ``dual periods''
$F_a = \int_{B_a} \Omega$ can be written as functions of $t$.
The auxiliary fields corresponding to $t^a$ can be written as:
\begin{eqnarray}
   D_{++}^a & = \int_{A^a} \left(\tilde{T} + \tilde{H}\right)
      \nonumber\\
   D_{+-}^a & = g_s \int_{A^a} \left(\tilde{F}-C^{(0)}\tilde{H}\right)
      \nonumber\\
   D_{--}^a & = \int_{A^a} \left(\tilde{T} - \tilde{H}\right)\ ,
\end{eqnarray}
where the tildes denote the projection of the forms into $H^{(2,1)}(X)$.
The auxiliary fields for $F_a$ are as above, only with $A^a$ 
replaced by $B_a$.
It is posssible to combine these statements
into a 'supermultiplet of three-forms' 
which incorporates all of the complex 
structure multiplets,
of the form
\be 
V = \Omega + \theta^i \theta^j D_{ij} + ... \ee
where $(i,j)$ run over $SU(2)$ doublet indices $\pm$.

A similar story holds for hypermultiplets in type IIA
compactifications.~\cite{Lawrence:2004zk}.  
The identification of auxiliary fields
for vector multiplets in IIA and hypermultiplets in IIB
is not yet completely understood.

\subsection{Soft SUSY breaking}

Vevs for auxiliary fields $D_{ij}$
break supersymmetry to $\CN=1$ or $\CN=0$.
For example, let $D_{--}^m \neq 0$.  
The SUSY transformations related to
$\hat{\theta}$ are broken, as
\be
   \delta \hat{\zeta}_\alpha = \hat{\epsilon}_\alpha D_{--}\ .
\ee
If in addition $D_{+-}= D_{++} = 0$, an $\CN=1$ SUSY is still unbroken.

When $D_{ij}$ is related by SUSY to the nonpropagating
complex structure deformations of $X$, 
one may fix its value by hand.  In this
case, SUSY is explicitly broken by couplings of
these nondynamical fields to the propagating modes.
One can show explicitly~\cite{Vafa:2000wi,Lawrence:2004zk} that
nontrivial vevs for $G = F - \tau H$, where 
$\tau= C^{(0)} + i/g_s$, breaks the supersymmetry 
generating translations along $\theta - i \hthet$,
and leads to a superpotential for complex structure 
moduli~\cite{GVW,Taylor:1999ii}.

We can also use fluxes to introduce soft SUSY breaking
terms in $\CN=1$ models with D-branes placed in CY backgrounds.
For example, let us study a D5-brane in type IIB
wrapping a rational curve $\CC$ inside $X$,
which preserves $\CN=1$ SUSY.  
Holomorphic deformations of $\CC$ correspond to
open string chiral multiplets,~\cite{Brunner:1999jq} with
superfield description $\Phi = \phi + \theta\psi + \theta^2 F_\phi$.  
To all orders in string perturbation theory, the superpotential
\be
   W = W(t^a, \Phi^i) = \sum_n g_n(t^a) \tr \Phi^n
\ee
for these modes depends only on the complex structure
moduli of $X$, and not on the K\"ahler class.
If $D_{ij} \neq 0$ is chosen so that the $\CN=1$ SUSY
preserved by the D5-brane is broken, one induces
explicit, computable SUSY-violating operators.  For example,
expand $t^a$ in the superspace direction
for which the D-brane preserves translation invariance:
$ V^a = t^a + \tilde{\theta}^2 F^a + ... $ where $F$ is the
corresponding auxiliary field.  The couplings $G_k$ should
be written as superfields, so that:
\be
   g_k \longrightarrow g_k(t^a)+\tilde{\theta}^2 F^a\p_a g_k\equiv 
   g_k +\theta^2\Delta_k \ ,
\ee
leading to soft SUSY-violating terms of the form
\be 
   \int d^2 \theta W + h.c. = \Delta_2 \tr \phi^2 + 
   \Delta_3 \tr \phi^3 + 
{\rm h.c.} + ....
\ee
such terms are induced in the presence of RR flux through
cycles whose periods appear in the functions 
$g_k(t^a)$.~\cite{Lawrence:2004zk}

\section{Connecting closed-string vacua by paths in
open-string field space}

In this section, we will show that by moving in 
open-string configuration space, it is possible to 
connect vacua with different 
values of closed-srting three-form flux.
This amplifies and applies some remarks
made in Ref.~\cite{AKV}.
%%about effects 
%of motion in open-string field space.  
%We
%begin by discussing 
%the case where the D-brane has a moduli 
%space.  
%In this case, 
%there is no 
%open-string modulus-dependent 
%superpotential, 
%and the jump in the flux is hard to detect.
%We then deform the CY
%-- in local models we can do this,
%these moduli are not dynamical -- 
%to obstruct the moduli of the curve,
%and make the effect visible.

Consider type IIB on a CY $X$, with a D5-brane wrapped on 
a holomorphic curve $\CC \subset X$ 
that is a member of a family $\CM$ of holomorphic curves
such that $\pi_1(\CM)$ is nontrivial.  Examples 
arise~\cite{kklmtwo,KMP} when $\CC$ is an 
exceptional curve in the resolution of 
an $A_1$ singularity over a Riemann 
surface $\CM=\CS_g$ of genus $g > 0$.

%Our result does not 
%depend on the existence of 
%this moduli space;
%we begin here to 
%provide control over the situation.
The moduli space of D5-branes is lifted by deforming the 
complex structure of $X$ in 
such a way that the family $\CM$ becomes
obstructed; such deformations are in correspondence~\cite{KMP}
%with holomorphic one-forms on $\CS_g$.  
with sections $dW_0 $ of the canonical bundle $T^\star \CS_g$.
After a generic such deformation, 
the moduli space is reduced to a  collection 
of isolated points where $0 = dW_0 $.
We will be discussing 
motion in off-shell configuration space
where $W'_0$ is not necessarily zero.
We may choose $W_0$ to be proportional
to some small control parameter $\epsilon$.
The potential hills between vacua are then parametrically 
small compared to the string scale, and the
field space for low-energy excitations
of the D5-brane is still well-described by $\CM$.

This correspondence between one-forms on the moduli
space and complex-structure moduli of the CY
implies a map from one-cycles of the 
moduli space $\CM$ to three-cycles 
of the CY.~\cite{Clemens,KMP}.
A path $\gamma$ maps to the three-cycle $\pi^{-1} \gamma$
obtained by fibering the exceptional curve $\CC_x$ 
over each point in $\gamma$.
Moving the D-brane around a 
loop $\gamma$ in the moduli space $\CM$ 
generates a quantum of  RR flux 
though the cycle $\pii \gamma$.~\cite{AKV}
This follows from the fact that 
the D5-brane is magnetically charged 
under the three-form flux.\footnote{
An illustrative analogy 
arises in Maxwell theory on $R^3 \times S^1$: start with
a magnetic monopole-antimonopole pair, and  
move the magnetic monopole around the circle. 
This causes the magnetic flux through the transverse plane 
to jump.
}
We can see this fact further manifest itself
in the superpotential.~\cite{lerche}

The effective superpotential governing the 
open-string moduli and complex structure moduli 
is:~\cite{Wittenobstruction,GVW,lerche}
\be W = W_{GVW} + W_{obstruction} 
=  \int_X \Omega \wedge G + \int_{\Xi} \Omega 
\ee
where $G$ is the three-form associated to
a linear combination of $D_{ij}$ preserving
the same SUSY as the D5-brane, and includes
the RR flux.~\cite{Lawrence:2004zk}
For deformations of a D5-brane from a rational
curve $\CC_0$ to a curve $\CC$, the 
three-chain $\Xi = \CC - \CC_0$. 
There are two ambiguities in 
defining this obstruction contribution
to the superpotential.~\cite{Wittenobstruction}
%One is in choosing the 3-chain $\Xi$ which bounds $\CC$.
%You can just as well add in some nontrivial 3-cycles.
%The second is in choosing a base point.
\begin{enumerate}
\item $\CC_0$ is a base point on the
closed string moduli space; changing $\CC_0$
changes $\Xi$ and so changes $W$ by an additive constant.
\item Since $H_3(X)$ is nontrivial, 
a 3-chain $\Xi$ such that
$\del \Xi = \CC - \CC_0$ is only determined 
up to the addition of an element of $H_3(X,{\bf Z})$.
This also additively changes the superpotential.
We will give a concrete example below.
\end{enumerate}
%Explain with the example.

Moving the brane in a loop $\gamma$ in
$\CM$ shifts the chain $\Xi$ 
by $ \Xi \mapsto \Xi + \pii \gamma$. 
This in turn shifts the superpotential 
by $ \int_{\pii \gamma} \Omega $.
By Poincar\'e duality,
this can be identified with 
\be \delta W = \int_X \Omega \wedge [\pii \gamma] .\ee
But this can be absorbed in $W_{GVW}$,
if the the RR flux shifts by $[\pii \gamma]$.
This possibility of interchanging 
contributions between the two terms in $W$ 
is made clearest by writing 
%$$ W = \int G \wedge \Omega + \int_{\Xi} \Omega 
%= \int ( G + [ \Xi ] ) \wedge \Omega 
%$$
%PD of chain? best to write
%$$ W = 
\be W =\int_{G + \Xi} \Omega .\ee
where $G$ is the 3-cycle Poincar\'e dual to the flux.

%When the brane has a moduli space, 
%the second term in the superpotential
%vanishes.  
%Next, we explain this effect in more detail in 
%a class of examples 
%where we can obstruct the moduli 
%of the curve.

% when you go around some loop in the moduli space,
%$\Xi$ gets some non-trivial 3-cycle added to it,
%and so the brane superpotential shifts by
%the corresponding period of the hol'c threeform.

%fluxes through chains: if they're 
%harmonic who cares.

Let us consider an explicit example of a patch of this model.
Consider the hypersurface in ${\bf C}^4$ given by
\be 
\label{definingeqn}
y^2 +u^2 +v^2 = W_0^\prime(x)^2 + f(x).\ee
In this example, we take $\CC$ to be an $S^2$
which can be resolved out of any double roots of the RHS of 
(\ref{definingeqn}) at a point $x$,
and $\Xi$ is this $S^2$ times a curve in the $x$-plane
ending at $x$.
There is no reason for $\Xi$ to be special Lagrangian,
and it is not.
%, and also that i'm here abusing the notation of the
%draft by referring to objects in the target space as $\CC$ rather than
%$\Phi(\CC)$ but you know what i mean.]
The superpotential is
$ W = \int_{\Xi} \Omega .$
All of the information about the threefold and its complex structure,
including this integral,
can be represented in terms of information on the Riemann surface 
$\Sigma$ at $u=v=0$, defined by
$ y^2 = (W_0^\prime )^2 + f $.
$\Sigma$ is a double cover of the $x$-plane,
each of whose fibers represent 
a two-sphere homologous to $\CC$.
The superpotential integral can be represented as 
\be W(x) = \int_{x_0}^{x} y(\tilde x) d\tilde x \ee
where $\CC$ is the $S^2$ over the point $x$,
and $x_0$ specifies the base-point curve.

\begin{center}
  \epsfxsize=50mm
  \epsfbox{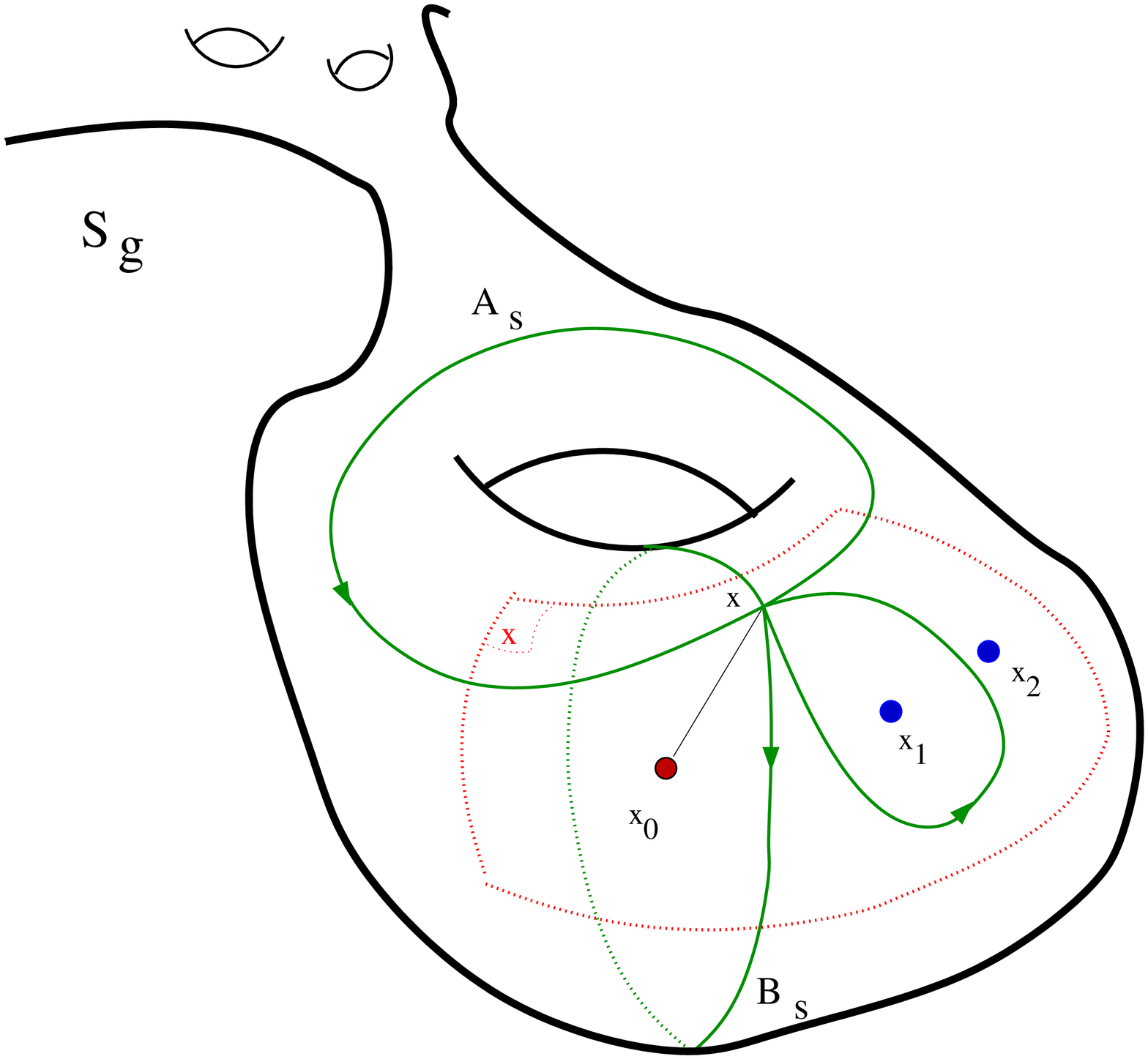}

{Fig. 1: {\em Each point in this picture represents a 
hemisphere
of the exceptional $P^1$ in the 
ALE singularity 
$ y^2 + u^2 + v^2 = 0 $.}}
\end{center}

It is important to distinguish $\Sigma$ 
from $\CS_g$.  The geometry of the $x$-plane embeds
into $\CS_g$ as shown in Fig. 1.
In this example, the moduli space of the curve $\CC$
when $W_0' = 0, f=0$ is $\CS_g$; 
for generic $W_0$ at $f = 0$, holomorphic curves appear 
only at the critical points of $W_0$.

%We can represent the elements of
%$H_3(X)$ apparent in (\ref{definingeqn})
%in terms of the $x$-plane.
%Three-cycles of the CY map to
%one-cycles of $S_g$, which map to
%contours negotiating cuts in the $x$-plane.
%For example, if we pick some particular
%pair of zeros of the RHS of 
%(\ref{definingeqn}),
%and we move the point $x$ around them along
%a contour $\gamma$
%then the superpotential clearly changes by
%\be W_0 \to W_0 + \oint_ \gamma y dx \ee
%which lifts back to the CY as
%\be \delta W = \int_{\pi^{-1}(\gamma) } \Omega
%= \int_X \Omega \wedge [\pi^{-1} (\gamma) ] \ee
%where $\pi$ is the projection which forgets about $u,v$
%and the brackets denote Poincar\'e dual.
%This is consistent with our claim
%that moving around a cycle $\gamma$ in the
%configuration space 
%$\Sigma$
%leads to a shift in the flux $G$ by
%\be G \to G + [\pi^{-1}(\gamma)] \ee

%At this point, let us dispense with superselection issues.
%This mechanism does not represent a dynamical instability
%of all flux vacua.
%Rather, it requires changing fields at infinity
%in $R^4$. 

The shift in the superpotential 
we have described is 
effected by changing the flux
through the cycle dual to the 3-cycle the
D5-brane sweeps out in moving through a loop in
$H_1(\CM)$.  
%As we stated, this is a consequence of
%the fact that the D5-brane is magnetically charged under
%the three-form field strength.  
We can see further that
this is consistent with rules for tadpole cancellation when
one turns on $N$ units of NS-NS three-form
flux $H$ through the three-cycle 
$\Xi$ that the D5-brane is sweeping out.

In this case, if the D5-brane sweeps out
a cycle, we have stated that it induces
a jump $\delta F = [ \pii \gamma ]$
in the RR flux.  Since $\int H \wedge F = M$
induces an RR 4-form tadpole that must be cancelled
by adding $M$ three-branes.
But this tadpole is precisely 
cancelled by D3-brane charge 
on the D5-brane which sweeps out the cycle
$\Xi$ with H-flux.  H-flux through $\Xi$ 
means that there is a gradient for
the B-field through the sphere the D5-brane is wrapping,
with respect to the direction on $\CM$ it is moving.
$H$-flux quantization means that the B-field will shift by
$2\pi N$ upon traversing the loop $\gamma$,
in units where $B = B + 2\pi$ when there is no brane.
Because of this, $B$ induces $N$ units of 
D3-charge via the worldvolume
Chern-Simons coupling $ \int_{D5} B \wedge C_{(4)}$.
This phenomenon is essentially identical to
the phenomenon described in~\cite{Kachru:2002gs}:
domain walls in ${\mathcal R}^4$ achieved by wrapping D5-branes
around cycles $\Xi$ with NS-flux $\int_{\Xi} H = N$
interpolate between vacua with D3-brane number differing by $N$.

%> it sounds like this effect should be something i know --
%> ok maybe here's a possibility
%> examples where the moduli space has handles aren't coming to me, so
%> suppose there's a singular point in the moduli space
%> at complex codim one [does this mean that the cycle itself needs to become
%> singular?]
%> then is it possible that in going around this singular point
%> there is some creation of *flux*,
%> by some integral [why integral?] 3-form $\delta G^{(3)} = \Gamma$?
%> then the superpotential would shift by
%> $$ \int_X \Omega\wedge \Gamma = \int_{[\Gamma] } \Omega $$
%> where $[\Gamma] \in H_3(X,Z)$
%> is the 3-cycle in question by which $\Xi$ shifted.

%energetics of  solitons where this happens 
%locally?  

This result implies that that in going around
what was apparently a loop
in this open-string moduli space $\CM$,
the string theory does not come back to itself.
Rather, the closed-string background is changed.
We conclude that the open-string 
moduli space in fact has no $\pi_1$.
For example, stable cosmic string solutions
%which one would have thought existed, 
corresponding to the putative loop do not exist.
One indication is that if one turns on the
obstruction superpotential, any loop in
${\mathcal R}^4$ 
for which the D-brane position loops around 
a cycle in $\CS_g$
will cross a domain 
wall.~\cite{Copeland:2003bj}  This
domain wall is a D5-brane wrapping the three-cycle
$\Xi$.  The jump in RR flux induced by
the motion in moduli space as one circles the
cosmic string is then cancelled by the
jump in flux induced by the domain wall,
so that the 4d solution is single-valued.
In fact this entire 
domain-wall-ending-on-cosmic-string
is one boundary-less D5-brane.

\begin{center}
  \epsfxsize=75mm
  \epsfbox{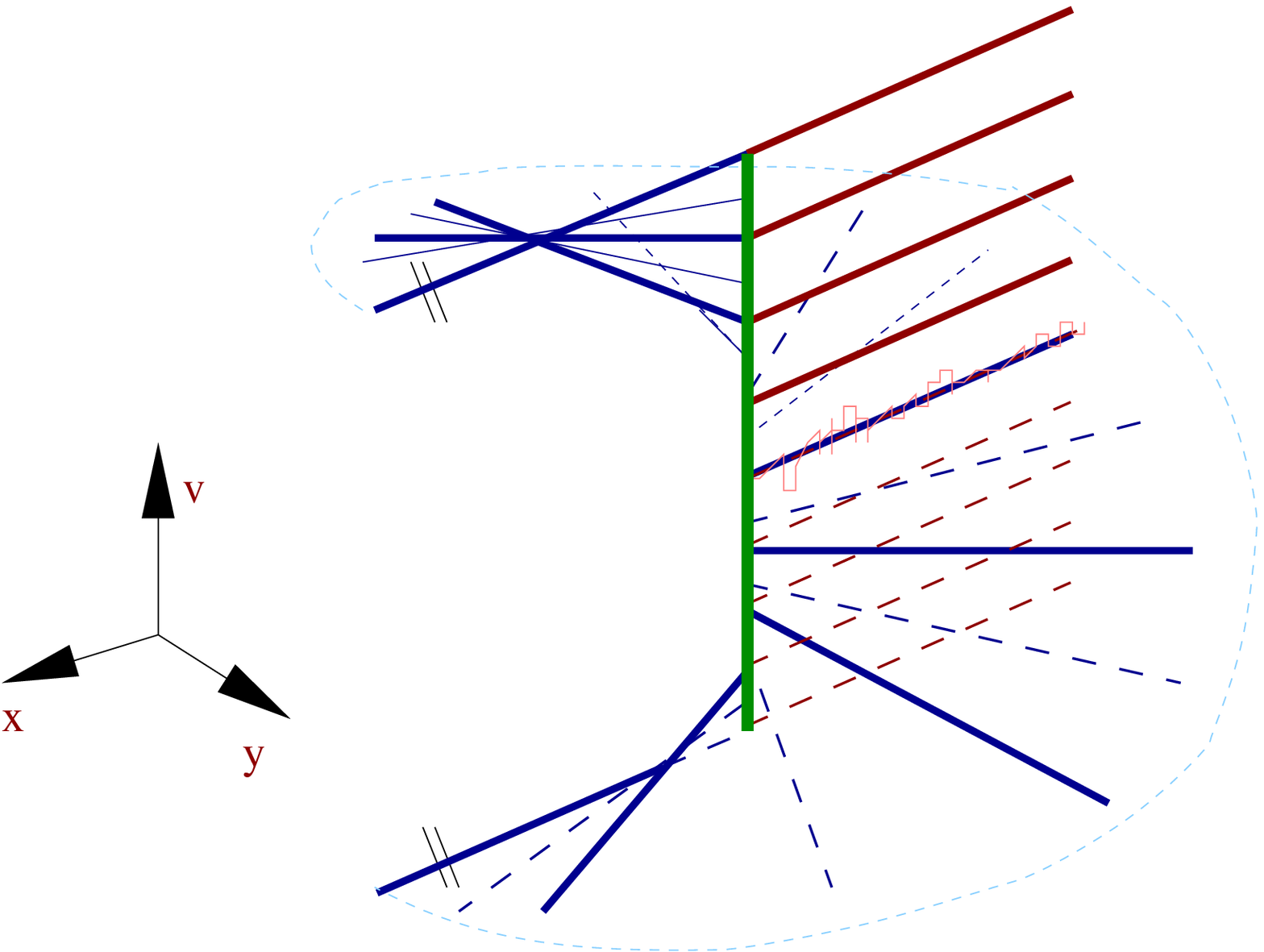}

{Fig. 2: {\em 
The vertical direction $v$ is the 
would-be cycle in the configuration space;
hashed lines are identified.
The blue lines represent the D5-brane on $\CC$
whose $v$ position depends on the argument of $x+iy$.
The red lines represent the domain wall D5-brane
wrapping $\Xi$.  
Note that their boundaries cancel.
This figure makes the instability of the configuration clear:
along the half-line where the brane crosses itself
(indicated by the wavy line),
it can annihilate; it can subsequently slip off the $v$-circle.
}}
\end{center}

Said another way, the moduli space
or low-energy field space is in fact a
{\em multiple cover} -- with infinitely many sheets -- 
of the Riemann surface $\CS_g$,
and the RR flux labels the sheets.
%This is an interesting kind of connectedness 
%[in the sense shamit likes to talk about]
%in the space of string theories. 
Note that this generically does involve going off-shell,
since in the presence of the obstruction superpotential
there is not in fact a moduli space.

A similar discussion implies that 
it is possible to interpolate between 
values of NSNS flux quanta 
by moving wrapped NS5-branes.

\section{$\CN=2$ Fayet-Iliopoulos terms in type II models}

Taylor and Vafa~\cite{Taylor:1999ii}\ showed that 
in local models of type IIB Calabi-Yau compactifications, the
superpotential~\cite{GVW}\ for complex structure moduli 
can arise from 
electric and magnetic Fayet-Iliopoulos (FI) terms~\cite{Antoniadis}
which spontaneously break the global $\CN=2$ SUSY to $\CN=1$.  
This is consistent with the 
identification of magnetic flux~\cite{Lawrence:2004zk,Vafa:2000wi}
with auxiliary fields, as the auiliary fields will be
equated to the FI terms on-shell.~\cite{Antoniadis}

One might ask whether the FI terms should be
identified as separate degrees of freedom, equated to
the magnetic flux via on-shell equations of motion.  This
appears to be the case.  Study a deformed conifold in type IIB with
vanishing 3-cycle $A$ and dual cycle $B$.  D3-branes
wrapping $A$ are light hypermultiplets charged with respect
to the vector multiplets associated with the period
$t_A$.  The hypermultiplet
can be written in terms of two $\CN=1$ chiral multiplets
$Q,\tilde{Q}$ with opposite $U(1)$ charge.  If we write
the $\CN=2$ supermultiplet $V = t_A + \ldots$ in terms
of an $\CN=1$ chiral multiplet $A$ and a $\CN=1$ vector multiplet,
the coupling of $Q,\tilde{Q}$ to $A$ includes the following
superpotential term:
\be
   W = \int d^2 \theta A Q \tilde{Q}
\ee
If the scalar component of $<Q\tilde{Q}>$ gets a vev, this
will appear as an electric FI term,
or a magnetic FI term with respect to the vector
multuplet associated to $t_B$.

It would be nice to show microscopically that such a vev induces
magnetic 3-form flux, and that the potential for $<Q\tilde{Q}>$
has discrete minima associated to different values of 
NS-NS and RR flux through $B$.  We can, however, note
that when $\int_A H = -K \gg 1, \int_B F = N \gg 1$,
this result is consistent with the conjectured field
theory dual.~\cite{Klebanov:2000hb,Kachru:2002gs}
This geometry is described by an 
$\CN=1$ $SU(NK + N)\times SU(N)$ gauge theory with
bifundamentals in $(NK+N,\bar{N})$ and $(\overline{NK+N}, N)$.
At low energies, the gauge invariant degrees of freedom include
``meson'' and ``baryon'' degrees of freedom, constructed
from the bifundamentals.  The mesons are dual to
motions of D3-branes in the Klebanov-Strassler geometry.
The baryons correspond to D3-branes wrapping the 3-cycles
of this geometry.~\cite{Klebanov:2000hb,Gubser:1998fp}
The space of vacua contains branches where
either the mesons or baryons have vevs.
Domain walls connecting the meson and baryon branches
were argued~\cite{Kachru:2002gs} to be 
dual to D5-branes wrapping $A$.
The disappearance of the D3-branes is consistent with
the tadpole cancellation arguments reviewed above.

%%%%%%%%%%%%%%%%%%%%%%%%%%%%%%%%%%%%%%%%%%%%%%%%%%%%%%%%%%%%%
% Doing Acknowledgement                                     %
%%%%%%%%%%%%%%%%%%%%%%%%%%%%%%%%%%%%%%%%%%%%%%%%%%%%%%%%%%%%%

\section*{Acknowledgments}

A.L. would like to thank the organizers
of QTS3 for organizing
an exciting conference, and for inviting him 
to speak there.  
We would like to thank Oliver DeWolfe 
for helpful comments and for the name of the second figure.
A.L. is supported in part by the NSF
grant PHY-0331516; J.M. is supported in part by
a Princeton University Dicke Fellowship and
in part by the DOE under Grant No. DE-FG03-92ER40701.

%%%%%%%%%%%%%%%%%%%%%%%%%%%%%%%%%%%%%%%%%%%%%%%%%%%%%%%%%%%%%
% Doing Appendix(ices)                                      %
%%%%%%%%%%%%%%%%%%%%%%%%%%%%%%%%%%%%%%%%%%%%%%%%%%%%%%%%%%%%%

%\appendix

%\section{HEADING FOR APPENDIX A}

%\renewcommand{\theequation}{A.\arabic{equation}}

%TYPE TEXT FOR APPENDIX A HERE.

%\section{HEADING FOR APPENDIX B}

%\renewcommand{\theequation}{B.\arabic{equation}}

%TYPE TEXT FOR APPENDIX B HERE.

%%%%%%%%%%%%%%%%%%%%%%%%%%%%%%%%%%%%%%%%%%%%%%%%%%%%%%%%%%%%%
% Doing references:                                         %
%%%%%%%%%%%%%%%%%%%%%%%%%%%%%%%%%%%%%%%%%%%%%%%%%%%%%%%%%%%%%

\end{document}